\documentstyle[prb,aps]{revtex}
\input epsfig.sty

\begin{document}

\advance\textheight by 0.2in
\twocolumn[\hsize\textwidth\columnwidth\hsize\csname@twocolumnfalse%
\endcsname

\draft
\begin{flushright}
{\tt to appear in Phys. Rev. B }
\end{flushright}

\title{Dynamics near the Surface Reconstruction of  W(100)}
\author{M.R. Baldan}
\address{Laborat\'orio Associado de Sensores e Materiais, 
Instituto Nacional de Pesquisas Espaciais, \\
12201-190 S\~{a}o Jos\'e dos Campos, S\~ao Paulo, Brazil \\
and Faculdade de Engenharia, Universidade S\~ao Francisco,\\
Itatiba, S\~ao Paulo, Brazil}
\author{E. Granato}
\address{Laborat\'orio Associado de Sensores e Materiais, 
Instituto Nacional de Pesquisas Espaciais, \\
12201-190 S\~{a}o Jos\'e dos Campos, S\~ao Paulo, Brazil}
\author{S.C. Ying}
\address{Department of Physics, 
Brown University, \\
Providence, Rhode Island 02912}
\maketitle

\begin{abstract}
Using Brownian molecular dynamics simulation, we study the surface
dynamics near the reconstruction transition of
W(100) via a model Hamiltonian. Results for the softening and broadening 
of the surface phonon
spectrum near the transition are compared with previous calculations and
with He atom scattering data. From the critical behavior of the
central peak in the dynamical structure factor, we also estimate the exponent of the
power law anomaly for adatom diffusion near the transition temperature.
\end{abstract}


]

The (1x1) to c(2x2) structural transition on W(100) has been one of the best
characterized and studied surface phase transitions \cite{ernst,thompson}. 
Besides the obvious interest 
in surface science as an important surface structural  transition, it provides a 
physical realization of the universality class of two dimensional XY model with 
four-fold symmetry breaking fields \cite{jose,ngkky}. Moreover, the  strength of the 
symmetry breaking field 
can be tuned by varying the coverage of adsorbates such as Hydrogen atoms. In particular, 
the field can be made to vanish at a particular coverage
to realize the pure two-dimensional XY model \cite{ngkky}. As such, the study and 
understanding of this transition has significance far beyond this system itself. 
It is of importance in the general field of statics and dynamics of two dimensional phase 
transitions. The equilibrium properties of the W(100) transition are  well understood through 
extensive experimental and theoretical investigations. On the other hand, the critical dynamics
for this transition has received relatively little attention. Experimentally, there exists only
one study which probes the softening and damping of the phonon modes \cite{ernst} near the 
critical temperature $T_{c}$. However, the details of the  critical dynamics such as the 
emergence of the central peak  and the dynamical critical exponents have not been investigated.
Previous theoretical studies of the dynamics for this transition  consist of a
finite slab geometry Molecular Dynamics (MD) simulation study \cite{wang} and
an analytic continued fraction (CF) expansion method \cite{hys}. Both of
these approaches have limitations near $T_{c}$ either because of the small
size limit in MD or because of the truncation of the CF expansion after a
few steps. In particular, although the softening and overdamping of the vibrational modes were 
confirmed in these studies, both these studies were not able to determine the  dynamical and 
static  critical exponents which characterize the critical region quantitatively.

Recently, there are renewed interests in studying   the critical
dynamics of the W(100) surface reconstruction using  different experimental probes. 
One approach that has been proposed is a
time resolved laser and LEED spectroscopy which consists of an initial laser
pulse that heats up the system followed by synchronized electron beam pulse
that probes the subsequent temporal variation of the order parameter for the
c(2x2) phase \cite{thompson}. This is a difficult experiment and has not been attempted yet. 
More recently, Xiao and Altman \cite{expdiff} propose to probe directly the temperature dependence 
of the diffusion constant of adsorbed atoms on W(100) using the
degradation of the diffraction signal from a laser-written grating on the
adlayer. The interest in this study arises from the theoretical prediction that the adatom 
diffusion constant would vanish at the transition due to the coupling to the critical fluctuations 
in the substrate \cite{nhy}. This experiment is under way and the results should yield better 
understanding of critical fluctuations near $T_{c}$ for this surface.

In view of the ongoing experimental study and the inadequacies of the previous theoretical 
investigations, we have undertaken  a new study of  the surface dynamics near the W(100) 
reconstruction through the direct simulation of the Langevin dynamics of a
two-dimensional effective Hamiltonian which has been very successful in the
study of the equilibrium properties of this transition \cite{hy91}. The
approximate CF study of the dynamics is also based on this Hamiltonian \cite{hys}. 
Unlike the previous MD studies, the simplicity of this effective
Hamiltonian allows us to use much larger sizes which is important for
probing critical fluctuation effects. We also avoid the truncation approximations 
used in the analytic continued fraction study \cite{hys}. The only drawback of using a
two-dimensional effective Hamiltonian is that the long wavelength phonons,
which penetrates deeply into the bulk of the substrate, cannot be described
accurately within this approach. However, unlike the excitations near the
critical wavevector $q_{o}=(\pi /a,\pi /a)$, these long wavelength phonons
are scarcely affected by the critical fluctuations and contributes little to
the strong temperature dependence of dynamical quantities near $T_{c}$. 
As shown below, with the larger size of the present study, we are able to collapse all 
our data for the dynamical structure factor based on a general dynamical scaling analysis. 
This allows us to extract the static and dynamical critical exponents, as well as extending 
the validity of the results arbitrarily close to  $T_{c}$.

The clean W(100) surface is described by an effective Hamiltonian in dimensionless
form \cite{hy91}
\begin{eqnarray}
H &=&\sum_{i}{\Large \{}\frac{p_{i}^{2}}{2}+\frac{A}{2}u_{i}^{2}+\frac{B}{4}
u_{i}^{4} \cr
&&+8H_{4}u_{ix}^{2}u_{iy}^{2}+C_{1}(u_{i}\cdot u_{j}+u_{i}\cdot u_{k})
\label{hamilt} \\
&&+C_{2}[u_{ix}(u_{j^{\prime }y}-u_{k^{\prime }y})+u_{iy}(u_{j^{\prime
}x}-u_{k^{\prime }x})]{\Large \}}  \nonumber
\end{eqnarray}
where $p_{i}$ is the momentum vector and $u_{i}$ the displacement vector of
a W atom from the ith site, $R_{i}=a_{o}(l,m)$, with $l$ and $m$ integers,
and $\ j=(l+1,m)$, $k=(l,m+1)$, $\ j^{\prime }=(l+1,m+1)$ and $k^{\prime
}=(l-1,m+1)$. The dimensionless parameters used in the simulation are the
same as for model I in the previous study \cite{hy91}, $A=-10$, $B=40$, 
$C_{1}=3.75$, $C_{2}=C_{1}/2$ and $H_{4}=-1.85$. For this choice of parameters,
the ground state displacement is $u_{0}=1$, and the transition temperature
has been determined previously from Monte Carlo
simulations \cite{hy91} to be $T_{c}=2.11$. The scale factors of
distance $L_{s}$ and temperature $T_{s}$  are  then chosen so that
$u_{0}$ and $T_{c}$ agree with the
experimentally observed values, $0.2 \AA$   and $230 K$, respectively, 
and have the values $%
L_{s}=0.2 $ \AA , $T_{s}=109\ $K. This effective Hamiltonian has been very 
successful in accounting for the static critical phenomena such as the 
detailed temperature dependence and anisotropy of the intensity of the 
LEED diffraction pattern \cite{hy91} near $T_{c}$. 
This Hamiltonian belongs to the universality class of the XY model with 
four-fold symmetry breaking field 
represented by the  $H_4$ coupling term in the Hamiltonian 
\cite{jose,huy87,ngkky}. Thus, the static critical exponents are 
nonuniversal, depending on the value of the marginal field $H_4$. 
In the large coupling limit, it should correspond to the Ising universality class 
when $C_2=0$. It can also be related, by renormalization-group 
arguments, to the Ashkin-Teller model which consists of two coupled Ising
models and is known to have a critical
line with varying static critical exponents \cite{domb}. However, 
less is known about the dynamical critical behavior. It is not known for 
example whether the dynamical critical exponent $z$ is  dependent on the 
value  of the  four-fold symmetry breaking field $H_4$. The same question
arises for the critical line of the  Ashkin-Teller model \cite {at}. 
The time scale $t_{s}$ for  this model  Hamiltonian is not adjustable  
and is determined by the relation 
$t_{s}=(m_{s}L_{s}^{2}/k_{B}T_{s})^{1/2}=0.28$ ps where $m_{s}$ is the 
W atom mass. In the previous CF study \cite{hys}, it was demonstrated 
that the softening and damping of the surface phonons observed in the 
He scattering experiment \cite{ernst} are well described quantitatively 
without any other adjustable parameters by the results of this model up 
to the point where the soft modes become completely overdamped. 

To simulate the dynamical behavior of the system described by the 
Hamiltonian in Eq. (\ref{hamilt}) we use standard methods of Brownian Molecular 
Dynamics \cite{allen}. The equations of motion for the particle coordinates are given by
\begin{equation}
m\frac{d^{2}u_{i}}{dt^{2}}+m\eta_{o} \frac{du_{i}}{dt}=-\frac{\partial H}{%
\partial u_{i}}+f_{i}  \label{mdeq}
\end{equation}
where $f$ is a random force with each component $\alpha $ satisfying

\[
<f_{i\alpha }(t)f_{j\beta }(0)>=2mk_{B}T\ \ \eta _{o}\ \delta (t)\ \delta
_{ij}\ \delta _{\alpha \beta }
\]
and $\eta _{o}$ is a microscopic damping parameter representing the frictional
coupling of the effective two dimensional system to the heat bath consisting of the other 
subsurface layers. 
A $N\times N$ square lattice with periodic boundary conditions has been used in
the simulations. Most calculations were done with $N=32$ and $N=46$. The
equations of motion (\ref{mdeq}) are integrated numerically with a time step
of $\delta t=0.01t_{s}$ in reduced units of time $t_{s}$.  The  damping 
coefficient parameter $\eta_{0}$ cannot be determined from equilibrium quantities like the other 
parameters in the effective Hamiltonian. To determine $\eta_{0}$, we compare the width of the phonon
modes calculated from the present model in a regime  far from $T_{c}$ with the experimental data
\cite{ernst}. For $\eta _{o} \ge 0.05$, the noncritical phonon modes have widths 
substantially larger than the observed experimental value. This sets an upper bound for
$\eta _{o}$. In our present investigation, we set
$\eta_{o}=0.05$. Near the transition, the critical dynamics and damping are dominated by the 
interparticle interactions and the precise value of $\eta _{o}$ is immaterial.

The  dynamic properties of the system can be most conveniently studied by 
examining the dynamic structure factor $S_{\alpha \beta }({\bf q},\omega )$
defined as

\begin{equation}
S_{\alpha \beta }({\bf q},\omega )=\int e^{i\ \omega \ t\ }
<u_{\alpha }^{*}({\bf q},t) u_{\beta }({\bf q})> \ dt  \label{sw}
\end{equation}
where $\alpha ,\beta $ represent the components of the displacement. We
consider in this paper mainly the longitudinal $S_{L}(q,\omega )$ and transverse $%
S_{T}(q,\omega )$ components of the tensor $S_{\alpha \beta }({\bf q},\omega )$,
obtained from displacements $u_{\alpha }$ along the direction of the vector $%
{\bf q}$ and perpendicular to ${\bf q}$, respectively. For a given wave vector ${\bf q}$, 
the longitudinal (transverse) phonon mode shows up as a peak in 
$S_{L}({\bf q}, \omega)$ ($S_{T}({\bf q},\omega )$), 
while the width of the peak corresponds to the damping  of the mode.
In Figs. 1, 2, 3 and 4,  the behavior of the calculated dynamic structure factor as a
function of temperature near the critical temperature $T_{c}\sim 230K$ are shown.
As can be seen from the Figures, as one approaches $%
T_{c}$ from above, both the longitudinal and transverse modes near ${ {\bf q}}_{0}=(\pi/a, \pi /a)$ 
start to soften and broaden until they are completely overdamped at $ T \sim 1.5\ T_{c}$. As one
further approaches $T_{c}$, the dynamics is completely dominated by a strong
central peak. These results are  qualitatively similar to that obtained in the previous study through
the analytic CF expansion approach \cite{hys}. Moreover, the peak
position and width of the soft phonon modes in the two approaches are approximately the same  
as shown in Figs. 5 and 6. Thus, the  good agreement between the 
theoretical results for the soft modes and the experimental He atom scattering data
as demonstrated  previously  \cite{hys} remains true. However, as shown in Fig, 6, 
the  present results show a stronger
central peak relative to the soft phonon mode  near the critical point when compared with  
the CF calculation \cite{hys}. 
This is due to the fact that a truncation of the CF after a few steps
introduces effectively a finite size limitation, leading to an underestimate of the critical fluctuation effects.
This difference in the relative strength of the central peak could be important in such physical quantities as 
the anomalous temperature dependence of the diffusion constant as discussed below. It also shows that the 
apparent convergence of the CF approach after a few steps can be deceptive, and the quantitative behavior of 
the critical dynamics 
needs a long expansion of the CF which may render the analytical approach impractical.

We now  examine the critical properties more closely by focusing on the central peak. It follows 
from standard scaling arguments that   near $T_{c}$,
the dynamical structure factor should
satisfy the scaling form

\begin{equation}
N^d S(\tilde{q},\omega )=\xi ^{z+\gamma /\nu }g_{\pm }(\tilde{q} \xi,\omega \xi ^{z})\text{ ,}
\label{scal}
\end{equation}
where $\tilde{q}=|\bf q - \bf q_o|$, $g_{\pm }(\omega )$ is a scaling function, $\xi \propto
|T/T_{c}-1|^{-\nu }$ is the divergent correlation length and $d$ is the system
dimension.  We now focus on 
$S(\omega)\equiv S(\tilde{q}=0,\omega)$. The results for $S(\omega)$ from the present calculation for
various $\omega$ and T near $T_{c}$ are plotted in Fig. 7. Note that at ${\bf q}={\bf q}_0$ or $\tilde q=0$,
the longitudinal and transverse modes are identical and $S(\omega)$ is isotropic. According to the scaling form
 in Eq. (\ref{scal}), the data in a
scaling plot $S(\omega )\ |T/T_{c}-1|^{\nu \ z+\gamma }\ \times \ \omega \
|T/T_{c}-1|^{-z\ \nu \ }$ should collapse on to a single  curve above or 
below $T_{c}$ respectively if the
exponents $\nu \ z+\gamma $ and $z\nu $ and the critical temperature $T_{c}$
are chosen correctly. This scaling plot is shown in Fig. 8 for a system size $N=32$. 
By adjusting the parameters for the exponents and $T_{c}$,
we have verified that the data for $S(\omega)$ indeed obeys the scaling hypothesis.
We obtain the values $T_{c}=2.07(4)$, $\nu z=1.9(4)$ and $%
\nu z+\gamma =3.8(4)$. The same analysis was performed for a larger system
size $N=46$ with the same results indicating that finite-size corrections are
negligible in the range of frequency and temperatures studied.
The resulting value of $\gamma =1.9(4)$ is consistent
with previous estimate from equilibrium simulations which found \cite{hy91} $%
\gamma =1.5(2)$ and $\nu =0.83(1)$. Combining this value of $\nu $ with the
present results from dynamic scaling, we get an estimate for the dynamic
critical exponent $z=2.3(4)$. It is interesting to note that according to
the dynamical scaling arguments, S($\omega=0$) diverges
as $|T/T_{c}-1|^{-(\nu \ z+\gamma)}$ as one approaches $T_{c}$. In the analytic CF
approach however, a truncation of  the continued fraction for S($\omega=0$) at the second stage 
yields an approximate  divergent behavior of $|T/T_{c}-1|^{-2\gamma}$ for S($\omega=0$). 
For the present model,
these two exponents $\nu \ z+\gamma$  and $2 \gamma$ are indistinguishable 
within our numerical error. Another interesting observation is that the 
value of the dynamical critical exponent 
$z$ determined here is close to the corresponding value for the Ising model
which lies in the range $2.08$ to $2.24$ as obtained by numerical and 
renormalization-group calculations \cite{pru}. It raises the interesting 
question of whether the dynamic exponent can be universal even when the 
static critical exponents are known to be dependent on the value of the 
marginal field $H_4$. Further work on dynamics of the XY model with four-fold 
symmetry breaking using more efficient Monte-Carlo methods for larger size 
systems are underway \cite{emma}. They should help to clarify more the 
issue of dynamic universality class for this model. However, for the related 
Ashkin-Teller model, recent numerical results \cite{at} show that the 
dynamical exponent $z$  does  not vary significantly along the 
nonuniversal critical line  compared with the  error estimates and 
is also close to the results for the Ising model. Thus, we expect that
the present estimate of $z$ may be a good approximation for our model with
a weak dependence on the parameters of the model. 
   
While the soft modes and the central peak can be directly probed
experimentally with He scattering studies \cite{ernst}, it also influences
indirectly physical quantities such as adatom diffusion constant which
depends strongly on the coupling of the adatom to the substrate excitations.
As shown previously \cite{nhy}, the friction  $\eta $ acting on the adatom due
to the coupling to the substrate phonon excitations is related
to  the zero-frequency limit of the dynamical structure factor,

\[
\eta \propto \int d^{2}\tilde{q}\ S(\tilde{q},\omega =0)\ |W(\tilde{q})|^{2}\ ,
\]
where $W(\tilde{q})$ denotes the Fourier coefficients of the coupling 
of the adatom to the substrate. Thus, adatom diffusion measurements \cite
{expdiff} will probe  the dynamic correlation function $S(\tilde{q},\omega =0)$
which diverges as the temperature approaches $T_{c}$ because of the central
peak excitations. According to the general scaling relation given in Eq. (\ref{scal}),
the friction coefficient $\eta$ should  diverge near the transition as 
$|T/T_{c}-1]^{-x}$, with the critical exponent \cite{expon} $x=\nu(z-2)+\gamma$.
The difference of the exponent for $\eta$ from that of  $S(\omega=0)$ arises from the 
integration over the wave vectors for the friction coefficient. Since $D \propto 1/\eta$, 
this leads to the prediction of a vanishing diffusion constant as one approaches $T_c$, with a 
power-law anomaly $D\propto |T/T_c -1|^x $. The divergent behavior of the friction parameter 
near $T_c$ is not just specific to this system. Diffusion anomalies near a surface  
preroughening transition has also been predicted \cite{tosatti} based on similar considerations. 
Finally, we have shown earlier \cite{ngy} that, the effective pinning strength of flux lattice 
in disordered type II superconductors should diverge at the flux-lattice melting transition  due to the 
coupling of the pinning center to the critical fluctuations. 
This could be the possible origin for the "peak effect" observed in high-$T_c$ superconductors 
in a magnetic field.  Recent experimental data \cite{ling}  show a precise correlation of 
the "peak effect" with the  vortex lattice melting transition, lending  support to this picture.

In conclusion, we have performed a Langevin simulation study of the 
dynamics of the W(100) 
surface reconstruction near the transition point. We are able to 
avoid the severe size limitations of the previous molecular 
dynamical studies by studying an effective two dimensional Hamiltonian and analyse the 
results through dynamical scaling theory.  
We have determined both the static and dynamical critical exponents 
for this transition which belongs to the universality class of XY model 
with four fold anisotropy. Finally, we have related the central peak 
of the phonon spectrum to the friction experienced by an adatom on 
this surface. The adatom diffusion constant on this surface is predicted to vanish at the 
transition point due to the divergent frictional damping. The comparison of our new results 
with forthcoming  adatom diffusion data would  provide a detailed picture and further 
understanding of the critical dynamics near this intriguing surface 
reconstruction. Furthermore, the divergent nature of the friction near the transition is not 
just specific to this system but should manifest itself near other continuous phase transitions 
such as the surface preroughening transition and the vortex lattice melting transition in 
type II superconductors.

\medskip This work was supported in part by a joint NSF-CNPq grant 
and by FAPESP: grant no. 99/02532-0 (E.G) and 98/00977-2 (M.R.B).

\newpage
\begin{figure}[tbp]
\centering\epsfig{file=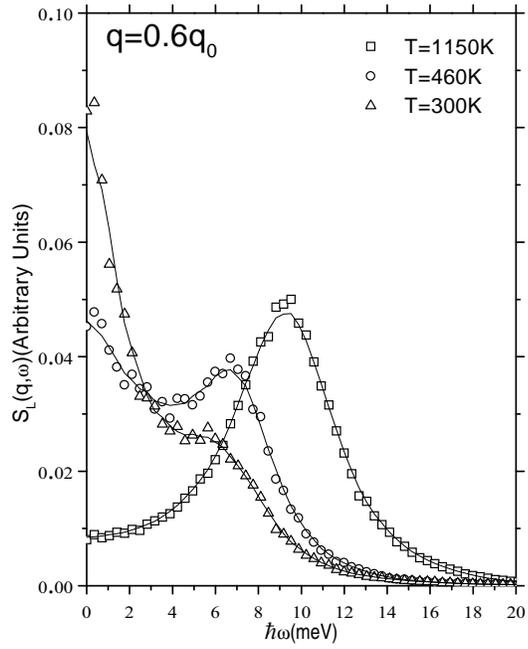,bbllx=5cm,bblly=0.6cm,bburx=17cm,
bbury=30cm,width=4.5cm}
\caption{Longitudinal structure factor $S_L({\bf q},\omega)$ for ${\bf q}=0.6 {\bf q}_o$ at different temperatures.}
\end{figure}

\newpage
\begin{figure}[tbp]
\centering\epsfig{file=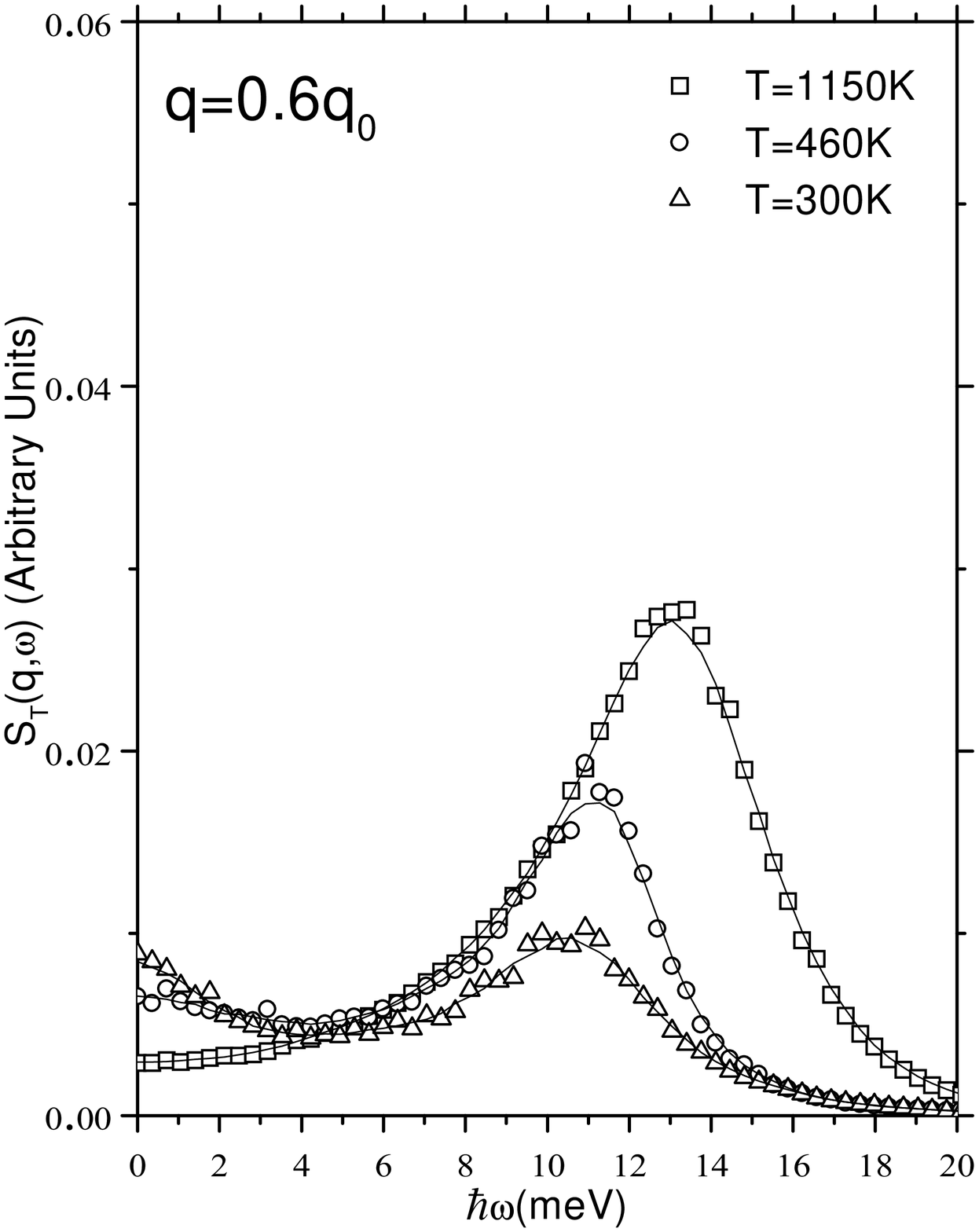,bbllx=5cm,bblly=0.6cm,bburx=17cm,
bbury=30cm,width=4.5cm}
\caption{Transverse  structure factor $S_T({\bf q},\omega)$ for ${\bf q}=0.6 {\bf q}_o$ at different temperatures.}
\end{figure}

\newpage
\begin{figure}[tbp]
\centering\epsfig{file=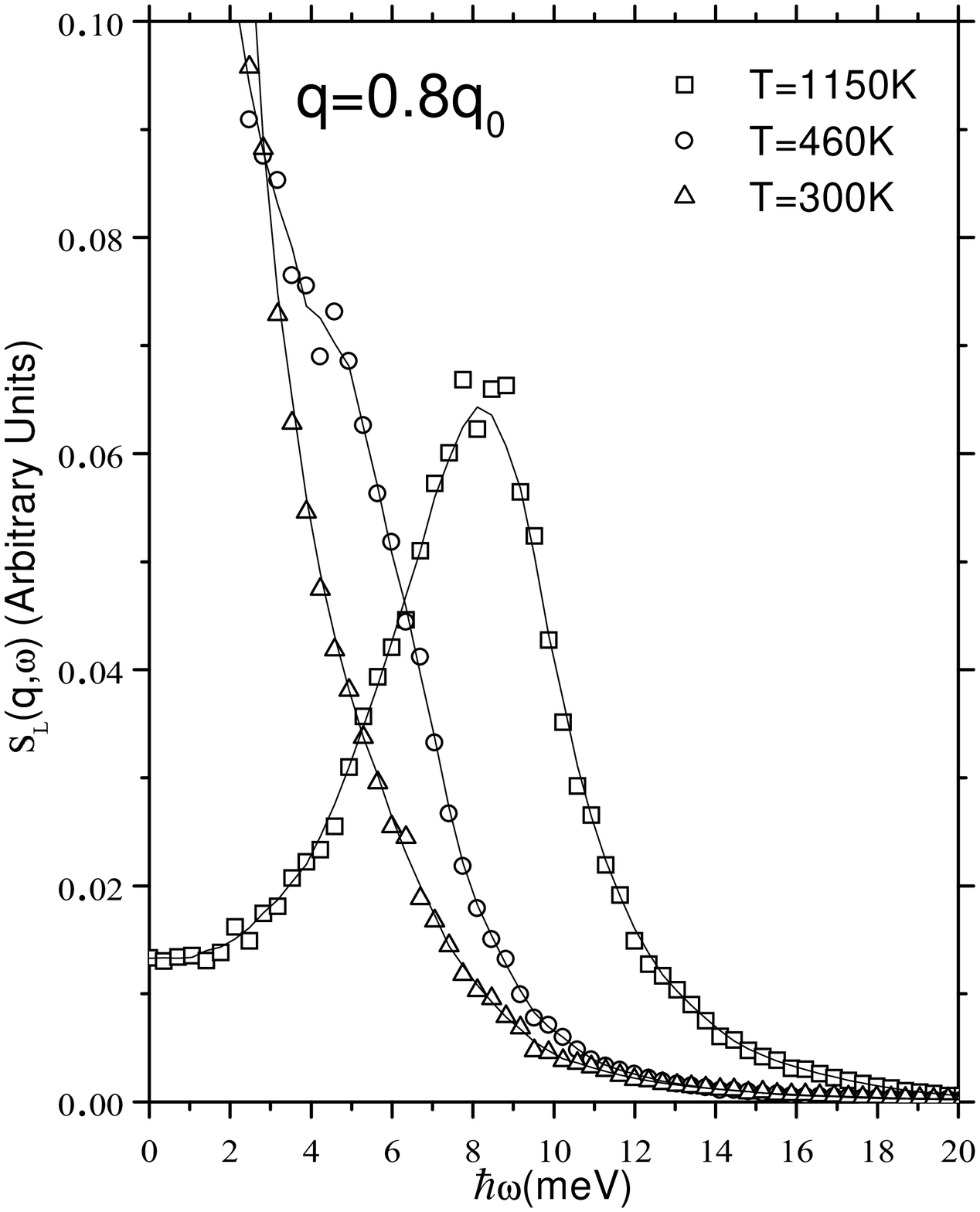,bbllx=5cm,bblly=0.6cm,bburx=17cm,
bbury=30cm,width=4.5cm}
\caption{Longitudinal structure factor $S_L({\bf q},\omega)$ for ${\bf q}=0.8 {\bf q}_o$ at different temperatures.}
\end{figure}

\newpage
\begin{figure}[tbp]
\centering\epsfig{file=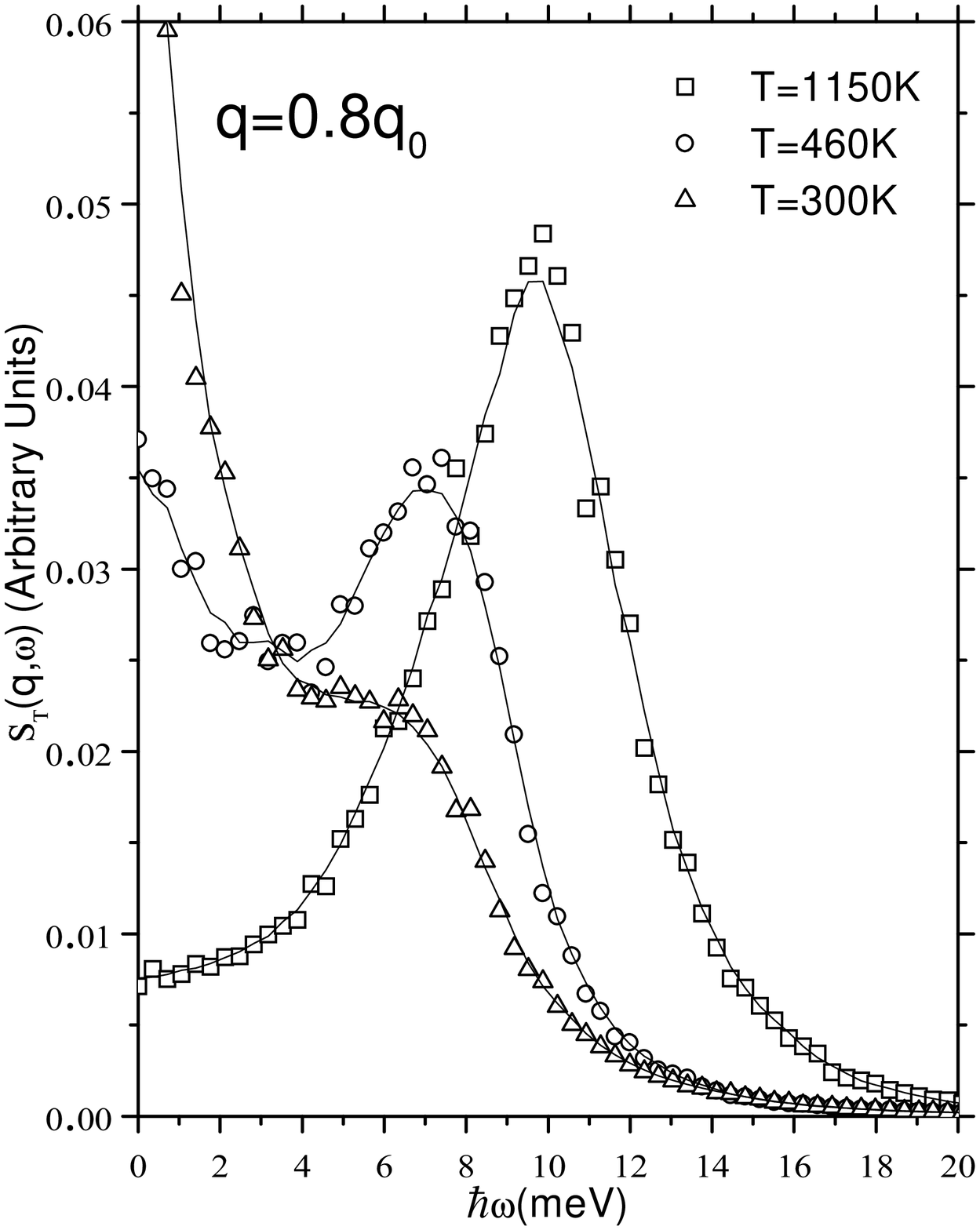,bbllx=5cm,bblly=0.6cm,bburx=17cm,
bbury=30cm,width=4.5cm}
\caption{Transverse structure factor 
$S_T({\bf q},\omega)$ for ${\bf q}=0.8 {\bf q}_o$ at different temperatures.}
\end{figure}

\newpage
\begin{figure}[tbp]
\centering\epsfig{file=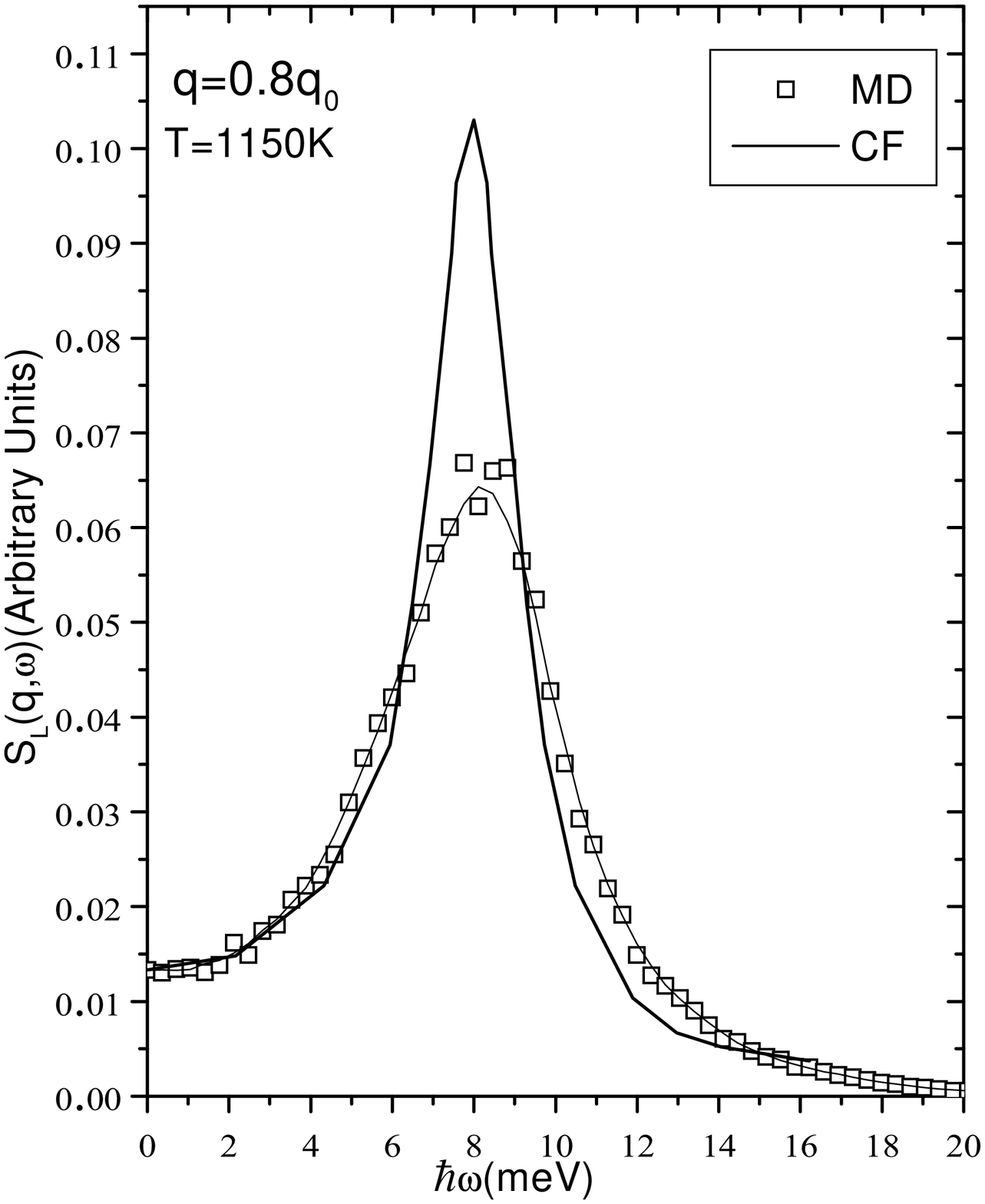,bbllx=5cm,bblly=0.6cm,bburx=17cm,
bbury=30cm,width=4.5cm}
\caption{Longitudinal structure factor $S_L({\bf q},\omega)$ far from 
$T_c$ for ${\bf q}=0.8 {\bf q}_o$ as obtained from 
the present study  and  the CF approximation \cite{hys}. }
\end{figure}

\newpage
\begin{figure}[tbp]
\centering\epsfig{file=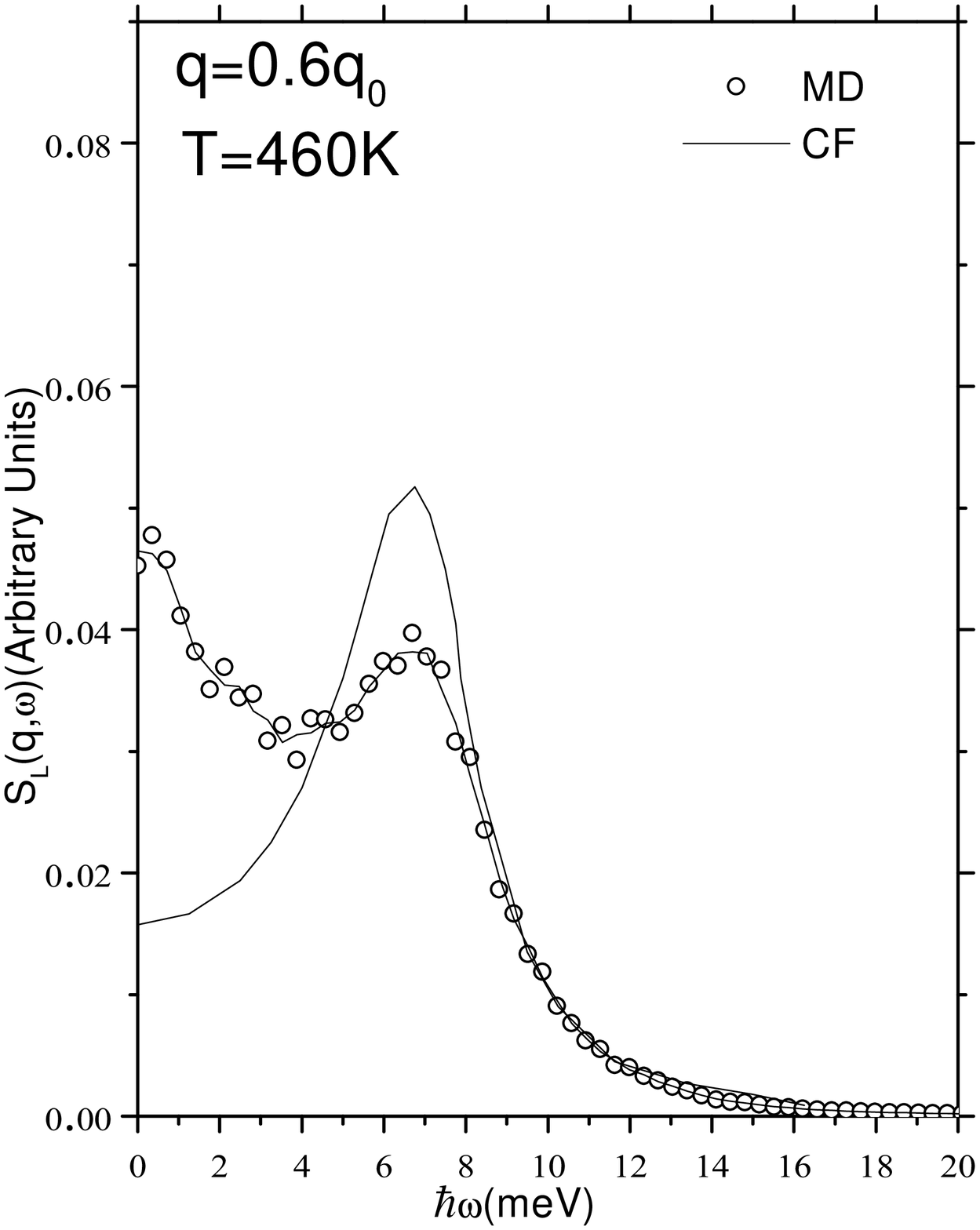,bbllx=5cm,bblly=0.6cm,bburx=17cm,
bbury=30cm,width=4.5cm}
\caption{Longitudinal structure factor $S_L({\bf q},\omega)$ close to $T_c$ for 
${\bf q}=0.6 {\bf q}_o$ as obtained 
from the present study and the  CF approximation \cite{hys}. }
\end{figure}

\newpage
\begin{figure}[tbp]
\centering\epsfig{file=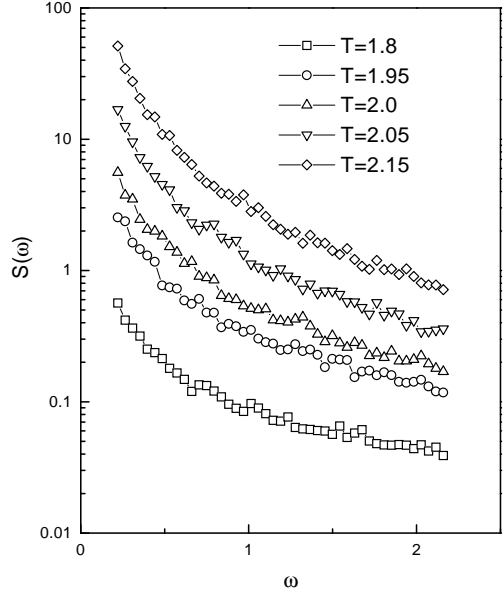,bbllx=5cm,bblly=0.6cm,bburx=17cm,
bbury=30cm,width=4.5cm}
\caption{Structure factor $S(\omega)$ at ${\bf q}={\bf q}_o$ as a function 
of dimensionless temperature $T$  and
frequency $\omega$  near $T_c$ . The units for $T$ and $\omega$ are $T_{s}$ and $t_{s}^{-1}$ as described in the text.}
\end{figure}

\newpage
\begin{figure}[tbp]
\centering\epsfig{file=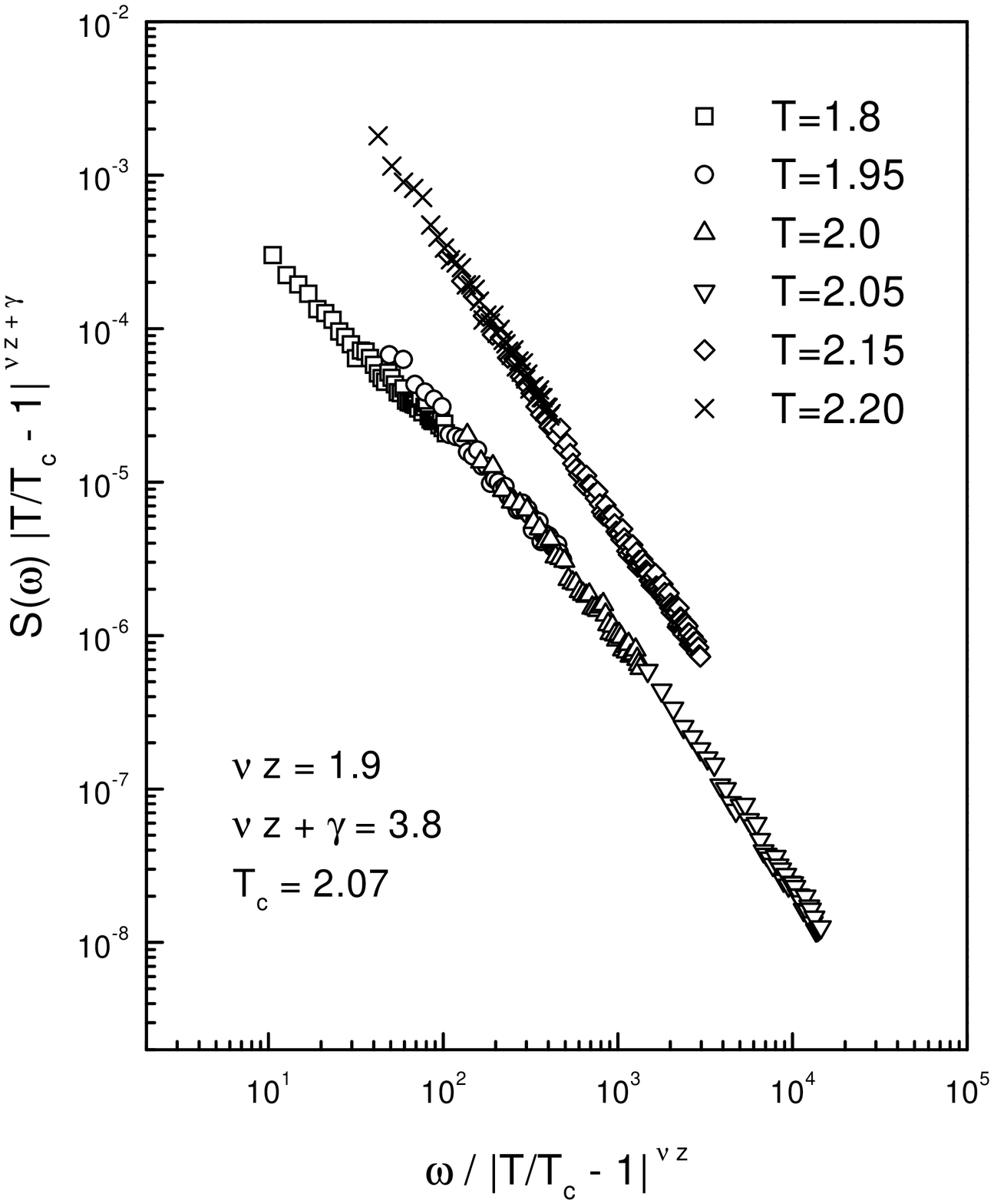,bbllx=5cm,bblly=0.6cm,bburx=17cm,
bbury=30cm,width=4.5cm}
\caption{Scaling plot  of 
$S(\omega) |T/T_c -1|^{\nu z + \gamma }  \times  \omega |T/T_c-1|^{-\nu z}$ from the data
in Fig. 7 obtained by adjusting the parameters $T_c$, $\nu z$ and 
$\nu z + \gamma$. The upper set of collapsing data corresponds to $T>T_c$ and
the lower one to $T<T_c$.}
\end{figure}


\begin{references}

\bibitem{ernst}  H.-J. Ernst, E. Hulpke, and J.P. Toennies, Phys. Rev. B
{\bf 46}, 16081 (1992).

\bibitem{thompson}  J.R. Thompson, P.M. Weber, and P.J. Estrup, in {\it Time-Resolved
Electron and Diffraction}, Edited by P.M. Rentzeps (SPIE, San Diego, 1995), p. 113.
(1994).

\bibitem{jose} J.V. Jos\'e, L.P. Kadanoff, S. Kirkpatrick, and D.R. Nelson,
Phys. Rev. B {\bf 16}, 1217 (1977).

\bibitem{ngkky} T. Ala-Nissila, E. Granato, K. Kankaala, J.M. Kosterlitz, and
S.C. Ying, Phys. Rev. B {\bf 50}, 12692 (1994).


\bibitem{wang}  C.Z. Wang, A. Fasolino, and E. Tosatti, Phys. Rev. B {\bf 37}, 2116 (1988).

\bibitem{hys}  W.H. Han, S.C. Ying, and D. Sahu, Phys. Rev. B {\bf 41}, 4403
(1990).

\bibitem{expdiff} X.D. Xiao and M. Altman, private communication.

\bibitem{nhy}  T. Ala-Nissila, W.K. Han, and S.C. Ying, Phys. Rev. Lett.
{\bf 68}, 1866 (1992).

\bibitem{hy91}  W.K. Han and S.C. Ying, Phys. Rev. B {\bf 41}, 4403 (1990).

\bibitem{huy87} G.Y. Hu and S.C. Ying, Physica {\bf 140A}, 585 (1987).

\bibitem{domb} B. Nienhuis, in {\it Phase transitions and critical 
phenomena},  edited by  C. Domb and J.L. Lebowitz, 
(Academic Press, London, 1987), Vol. 11, p. 1. 

\bibitem{at} Z.B. Li, X.W. Liu, L. Sch\"ulke, B. Zheng, Physica {\bf A245},
485 (1997).

\bibitem{allen}  M.P. Allen and D.J. Tildesley, {\it Computer Simulation of
Liquids} (Oxford University Press, 1987)

\bibitem{pru} V.V. Prudnikov, A.V. Ivanov, and A.A. Fedorenko, JETP Lett. {\bf 66}, 635 (1997).

\bibitem{emma} E. Falck and T. Ala-Nissila, private communications

\bibitem{expon} The expression for the critical exponent $x$ obtained here 
from dynamical scaling is more general than the one obtained previously
\cite{nhy} based on an approximate relation between $S(q,\omega)$ and
the displacement susceptibility.

\bibitem{tosatti} P. Prestipino, G. Santoro, and E. Tosatti, Phys. Rev. Lett. {\bf 75}, 4468 (1995).

\bibitem{ngy}  T. Ala-Nissila, E. Granato, and S.C. Ying, J. Phys: Condens. Matter {\bf 2},
8537 (1990).

\bibitem{ling} J. Shi, X.S. Ling, R. Liang, D.A. Bonn, and W.N. Hardy, Phys. Rev. B {\bf 60}, R12593 (1999).

\end{references}
\end{document}